\documentclass{aa}
\usepackage{txfonts}
\usepackage{natbib}
\usepackage{graphicx}
\usepackage{lscape}
\usepackage{booktabs}
\usepackage{placeins}
\begin{document}

   \title{The L1157 filament as traced in HNCO by the LMT
}

   \author{A. I. G\'omez-Ruiz
          \inst{1,2},
          C. Codella\inst{3}, O. S. Rojas-Garc\'ia\inst{1}, R. Bachiller\inst{4}, D.  S\'anchez-Arg\"uelles\inst{1,2}, M. Chavez Dagostino\inst{1}, G. Narayanan\inst{5}, A. Batta\inst{1,2}, V. G\'omez-Rivera\inst{1}, I. Rodr\'iguez-Montoya\inst{1,2}, E. E. Mart\'inez-Garc\'ia\inst{1,2}, R. Naranjo-Romero\inst{1}.
          }
     \institute{Instituto Nacional de Astrof\'isica, \'Optica y Electr\'onica, Luis E. Erro 1, C.P. 72840, Tonantzintla, Puebla, M\'exico \\
\email{aigomez@inaoe.mx}
         \and
         Secretaría de Ciencia, Humanidades, Tecnología e Innovación, Av. Insurgentes Sur 1582, C.P. 03940, Ciudad de México
         \and
   INAF-Osservatorio Astrofisico di Arcetri,
              Largo Enrico Fermi 5, I-50125 Firenze, Italy
        \and
   Observatorio Astron\'omico Nacional (OAN),
              Alfonso XII, 3, E-28014, Madrid, Spain
        \and
   Department of Astronomy, University of Massachusetts, Amherst, MA 01003, USA
             }

   \date{Received ; accepted }

  \abstract
   {The L1157 outflow is a prototype of chemically active outflows where to look for molecular species linked to complex chemistry.}
   {With the aim of imaging the large (parsec) scale structure of key prebiotic molecules, we have mapped the whole blue-lobe of the L1157 outflow with the Large Millimeter Telescope (LMT), proving the potential of this facility to detect weak and extended molecular emission.}
   {Isocyanic acid, HNCO, an important precursor of complex chemistry, was chosen to be observed in its (4$_{0,4}$--3$_ {0,3}$) transition at 87.925 GHz with the SEQUOIA focal-plane array, covering an area of 3$\times$3 arcmin$^2$ centered at the position of the main shock L1157-B1, which included a large portion of the vicinity of the protostellar region.}
   {The HNCO emission was detected not only from the outflow but it also was found tracing a structure of about 0.3 pc in length, corresponding to the filament from which the protostellar system was born.}
   {The HNCO molecule was found to be a tracer of the densest part of filaments in molecular clouds, and given its importance in the formation of more complex molecules, its presence in the cold, quiescent filamentary gas indicates that the building blocks of complex chemistry are already well established in stages prior to protostellar warming or shocks. By allowing the detection of structures related with the filament, these observations proved the LMT potentialities to map weak and extended components in molecular clouds.}

\titlerunning{iCOMs L1157}
\authorrunning{G\'omez-Ruiz, codella, et al.}

   \maketitle
\nolinenumbers

\section{Introduction}

\begin{figure*}[!ht]
\centering
\includegraphics[angle=0,width=0.8\textwidth]{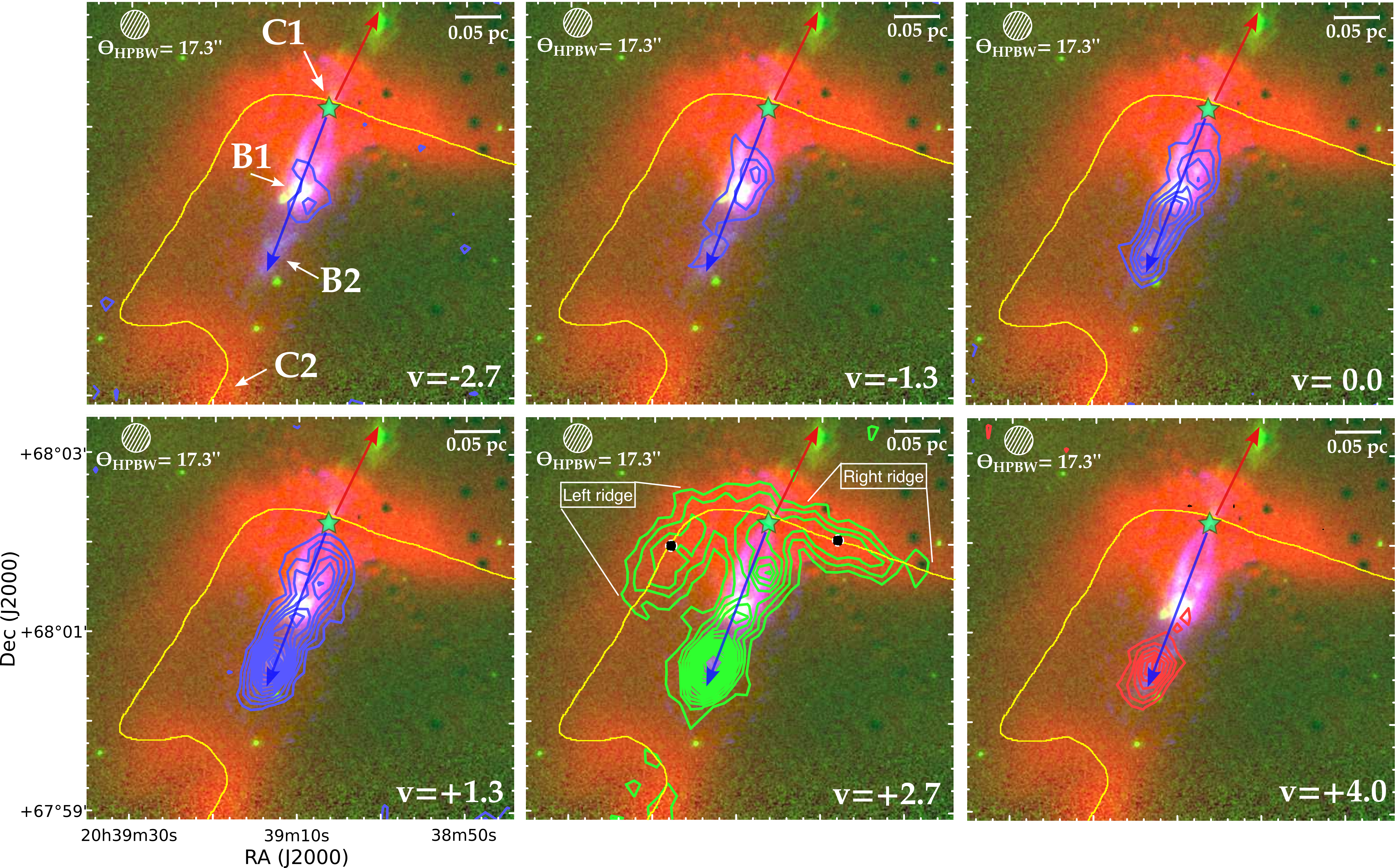}
\caption{Channel maps of the HNCO (4$_{0,4}$--3$_{0,3}$) emission from the southern part of the L1157 outflow overlaid on a composite image consisting of the 250 micron emission from \cite{Sharma2020} (red), 8 micron emission from Spitzer (green) and the CO (1--0) integrated emission from \cite{Gueth1996} (blue). Contour maps have values of 3, 5, 7, 9, 11, 15, 20, 25, and 30, times the rms of 9.3 mK (Ta$^*$); showed in blue, green and red, to indicate blue-shifed emission, cloud velocity emission, and red-shifted emission, respectively. The LSR velocity (in km s$^{-1}$) is indicated at the lower-right (with a channel resolution of $\approx 1.33\mathrm{~km\,s^{-1}}$, panel labels rounded to one decimal place imply that channel central velocities display alternating increments of $1.3$ and $1.4\mathrm{~km\,s^{-1}}$). The stared symbol indicates the position of the driving source L1157-mm; the main shocks are labeled B1 and B2,  while C1 and C2 are the far-IR clumps. The yellow line shows the approximate location of the filamentary structure as proposed by \cite{Sharma2020} and the outflow lobes are represented by the arrows. At the cloud velocity we highlight the parts of the HNCO emission that we propose are part of the filament, and that we call left and right ridge. We also indicate with black squares the position on which we have taken some of the sample spectra shown in Figure 2.}
\label{channels}
\end{figure*}

The current view of stellar birth considers the formation of filaments as a natural consequence of supersonic turbulence during cloud collapse, and that the processes leading to filament shaping and its evolution set the initial conditions for star formation \citep[for recent reviews, see][]{Hacar2023,Pineda2023}. Observations in the far-infrared, sub-mm/mm and radio waves have revealed a widespread presence of filaments in the interstellar medium, with the latest census counting nearly 23 000 filamentary structures in our own Galaxy, covering a wide range of scales \citep{Hacar2023}. Typical tracers of the filamentary structure are the submm/mm continuum, the HI line, and rotational transitions of molecules such as $^{13}$CO, C$^{18}$O, HCN, and N$_2$H$^+$, the later of which tracing preferentially denser gas \citep{Priestley2023}. 

L1157 is part of a complex of molecular clouds named L1147/L1158 located in Cepheus at estimated distances in the range of $\sim$ 330-370 pc  (352$\pm$18 pc reported by \citet{Zucker2019}, while 340$\pm$3 pc was determined by \citet{Sharma2020}). Large-scale maps of the far-infrared emission show that the coma-shaped dark cloud named LDN 1157 breaks into a filamentary structure of about 1.2 pc in length and 0.09 pc in width, in which one half runs from east to west and the other half change to a north-south direction \citep[][see figure \ref{channels}]{Sharma2020}. Two high-density clumps are identified within the filament, named C1 and C2 \citep[with peak values of dust column densities of 1.3$\times$10$^{22}$ and 9$\times$10$^{21}$ cm$^{-2}$, respectively, according to][]{Sharma2020}. Almost at the eastern end of the east-west segment lies the C1 clump that harbors the L1157-mm protostellar object that has recently been revealed as a binary system \citep{Tobin2022}. At the southern end of the north-south segment and beyond the L1157-mm blue-lobe outflow, resides the C2 clump, which is less dense than C1 and, given the absence of detectable 70$\mu$m emission, it is thought to be a starless core.

The L1157-mm protostellar system drives a precessing and episodic jet of matter at supersonic velocity \citep{Gueth1996,Zhang2000,Nisini2010,Podio2016}. Most of the studies have focused on the bright blue shifted southern outflow lobe, in which the jet has excavated two main cavities, the apices of which are called B1 and B2. Located at the apex of the more recent cavity, the bright bow shock B1 has a kinematical age of $\sim$1000 years \citep{Podio2016} and its intense chemical activity has been revealed by numerous investigations \citep[e.g.][]{Bachiller2001,Arce2008,Yamaguchi2012,Benedettini2013,Burkhard2016,Lefloch2017,Lopez2024}. On the other hand, the B2 shock is thought to be an older event \citep[$\sim$1800 years:][]{Podio2016}, with less chemical diversity than B1 \citep{Bachiller2001}. This older shock, on the plane of the sky, is located close to the C2 clump, at its northern end. 

Millimeter continuum, $^{12}$CO, C$^{18}$O, and N$_2$H$^+$ observations have revealed the detailed physical and chemical characteristic of the L1157 filament and its cores \citep{Sharma2020}. The emission of $^{12}$ CO that traces diffuse and external gas shows double-peaked line profiles all along the filament. The C$^{18}$O peak velocities show a remarkable velocity coherence of the filament, with LSR values of 2.65$\pm$0.05 km s$^{-1}$. In contrast, the N$_2$H$^+$ molecule shows line profiles with peak LSR velocities in the range of 2.64 to 2.78 km s$^{-1}$, indicative of the internal kinematics of the cores, the velocities of which are 2.62 km s$^{-1}$(C1) and 2.72 km s$^{-1}$ (C2). In addition, interferometric observations have shown that the emission from C1 is coincident with the elongated absorption feature at 8$\mu$m that has been suggested to be the flattened envelope of the protostellar system \citep{Chiang10}. The millimeter continuum and N$_2$H$^+$ are stronger in C1 than in C2, and the opposite tendency is found for C$^{18}$O, which suggests that CO is more depleted in C1 than in C2, which in turn suggests that C2 is chemically younger than C1 \citep{Sharma2020}. 

In this letter, we leverage the unique capabilities of the Large Millimeter Telescope (LMT) to bridge the observational gap between high-resolution interferometric studies and earlier single-dish surveys. By utilizing the LMT, we sample the 3mm line emission with a beam approximately two times smaller than that of the IRAM 30m. The increased sensitivity and resolution allow us to recover the filament/envelope-scale emission that interferometers mostly miss, while providing the clarity needed to resolve intermediate-scale structures. As demonstrated here, this capability has been crucial for identifying the filamentary emission of the HNCO molecule, targeted in this study as a key precursor of complex organic chemistry, and underscores the LMT's important role in connecting disparate spatial scales in star formation regions.

\section{LMT observations}

The observations were carried out with the 50-m diameter Large Millimeter Telescope Alfonso Serrano \citep{Hughes2020} in 2022 (season 2021-S1) during the nights of May 4, 10, 14, 17, 18 and 24. We used the heterodyne 16-pixel array SEQUOIA \citep{Erickson99}, centered at the sky frequency of 87.85 GHz. The WARES spectrometer was used in the wide mode with a total bandwidth of 800 MHz and 2048 channels, corresponding to 2732 km s$^{-1}$bandwidth and velocity resolution of $\approx$ 1.33 km s$^{-1}$ at the center of the band. Therefore the spectral setting includes the HNCO (4$_{0,4}$--3$_{0,3}$) line at 87.925 GHz (E$_U=$ 10.5 K) targeted in this study, but also NH$_2$CHO (4$_{1,3}$--3$_{1,2}$) at 87.848 GHz (E$_U=$ 13.5 K) reported in Appendix A \citep[rest frequencies and upper energies obtained from CDMS:][]{Muller2005}. A series of maps of 180$''\times$180$''$ in size were made with the On-The-Fly (OTF) technique, centered at the position of the B1 shock of the L1157 outflow, but effectively covering the whole blue-lobe and the region around the driving protostelar system. The 225 GHz zenith opacity during the observing nights varied from 0.05 to 0.29, and the system temperatures measured with SEQUOIA were in the range of $\sim$80-116 K. Pointing accuracy was checked about every hour on IRC+70171, with offset values in the range of 0.5-4 arcsec. The data reduction included the identification and flagging of bad data from certain beams (beams 0, 1, 14 and 15 were excluded for the first two nights, while only 0 and 5 were excluded for the rest of the nights) and birdies along the frequency axis. In the gridding process the final data cube was convolved to a beam of 17.3$''$ (the nominal HPBW at HNCO rest frequency is 15.9$''$). The final rms of the cube is 7.1 mK per channel width. To convert antenna temperature to main beam temperature an efficiency of 65\% is used, according to the latest SEQUOIA commissioning data \citep{Hughes2020}.

\section{HNCO filamentary structure}

\begin{figure}[!ht]
\centering
\includegraphics[angle=0,width=7cm]{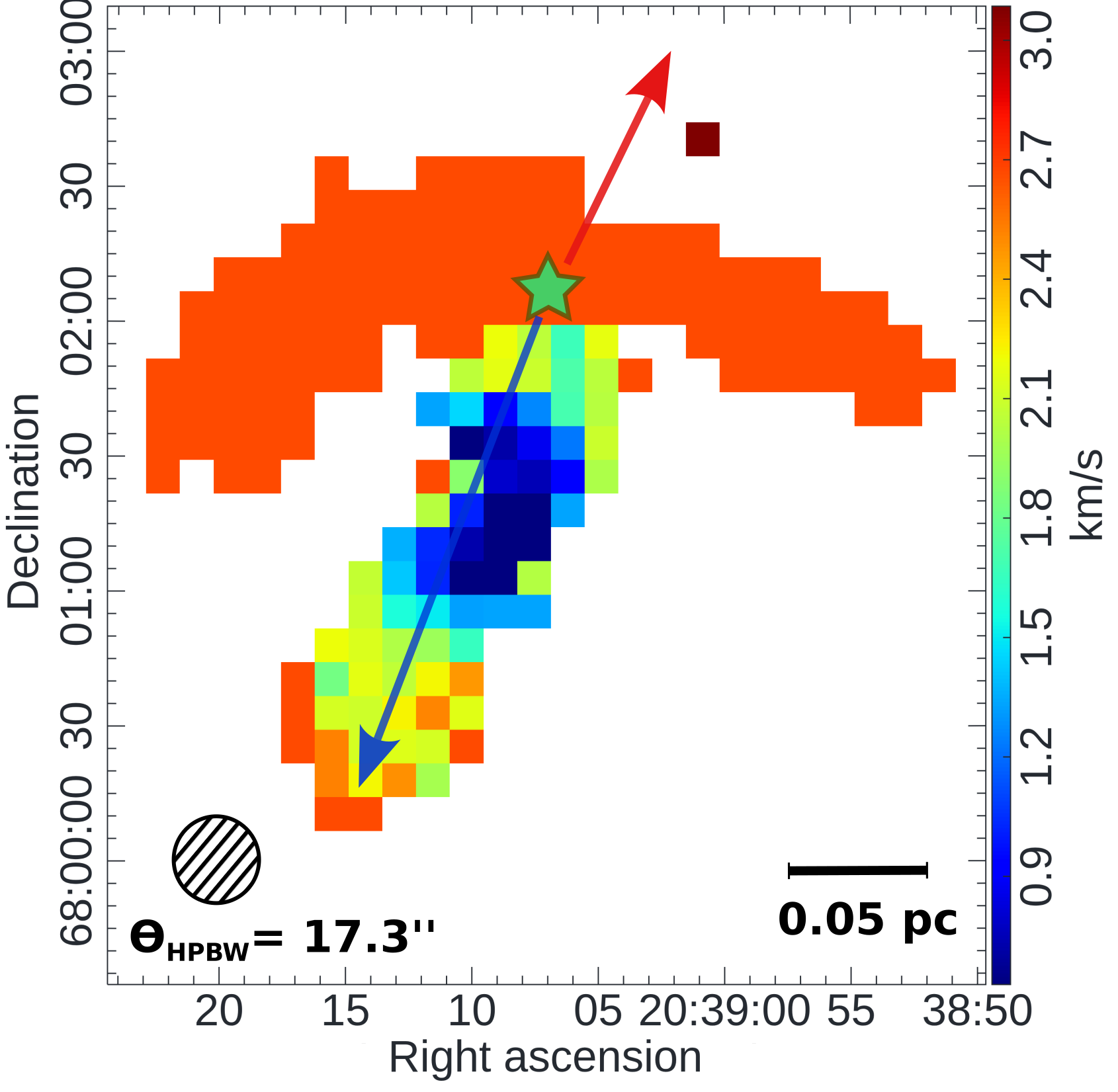}
\caption{Velocity field (moment 1) map of the HNCO (4$_{0,4}$--3$_{0,3}$) emission. The star symbol and the arrows indicate the protostellar position (L1157-mm) and the blue/red outflow lobes, respectively. The beam size and a physical scale bar are also shown. }
\label{moment1}
\end{figure}

In Figure \ref{channels} we show the channel maps of the HNCO (4$_{0,4}$--3$_{0,3}$) emission overlaid on a composite three color image consisting of 250 $\mu$m, 8$\mu$m and CO emission. It is clear that at velocities shifted from the cloud value ($\sim$2.6 km s$^{-1}$), the emission traces the blue lobe of the L1157 outflow, while the channel coincident with the cloud velocity (highlighted with green contours in Figure \ref{channels}) shows additional structures to the west and southeast of the position of the central protostellar region, that match the filamentary structure traced by the 250$\mu$m emission (seen in red in Figure 1). These additional HNCO structures are coincident with the so-called the east-west and north-south segments of the L1157 filament found by \citet{Sharma2020}. The highest velocity structures correspond to shocks already identified along the cavities that form the blue lobe, such as B1 and B2. Some parts of these additional structures seen only at the cloud velocity have not previously been reported in molecular emission, and to distinguish them from from the whole extension of the filamentary structure seen in the far infrared, we call them left ridge and right ridge in Figure 1 (see detailed description in the next paragraph). The velocity integrated emission is dominated by the strong outflow emission (see moment 0 map in Figure \ref{moments}). A clear picture of the velocity pattern traced by HNCO can be seen in the moment 1 map shown in Figure \ref{moment1}, in which the highest velocity blue-shifted emission is concentrated towards the B1 and B0 shocks, while the lowest velocity is tracing elongated structures to the west and southwest of the driving source(s) already mentioned. With a spectral resolution of $\sim$1.3 km s$^{-1}$, the emission from the filament and the envelope/core remains blended within a single channel ($V_{lsr} = 2.7$ km s$^{-1}$), preventing a clear kinematic decomposition of these structures in the moment 1 map (all moments maps are shown in Appendix A for reference). From gaussian fits to the left and right ridge structures, and using a distance of 340 pc, their sizes are in the range of 0.15-0.16 pc x 0.04-0.06 pc. 

Looking  at the N$_2$H$^+$ interferometric maps provided by \cite{Chiang10} (figure 2 in their paper), it seems that this molecule is tracing the same elongation to the west of L1157-mm that we see in HNCO (right ridge), but misses the southeast elongation (left ridge). The single-dish and interferometric N$_2$H$^+$ maps by \citet{Tobin2011} also show the western elongation (right ridge) but also emission coincident with the outflow cavity walls; however, it is noticeable that a hint of the southeastern elongation (left ridge) is found in their NH$_3$ (1,1) VLA map. Therefore HNCO and N$_2$H$^+$ are likely tracing the dense part of the filament, but the main difference is that while N$_2$H$^+$ is peaking towards the positions of L1157-mm (the center of clump C1), HNCO does not show an emission peak there, but rather presents what seems to be well defined large-scale clumps to each side of C1, the elongations to the west and east-southeast (right and left ridge structures, respectivey). The critical density of this transition ($\sim$10$^6$ cm$^{-3}$ at 20 K collisional temperature) indicates that it is likely tracing extended high-density gas. Unfortunately, a full comparison between the kinematics of $\text{N}_2\text{H}^+$ and $\text{HNCO}$ is limited by our spectral resolution of $\sim$ 1.3 km s$^{-1}$. This resolution is insufficient to resolve the velocity patterns of less than 1 km s$^{-1}$ show by \citet{Tobin2011} data within the structures we identify as the filament elongations. Nevertheless, it is noteworthy that the $\text{N}_2\text{H}^+$ FWHM line widths at the ridge structure positions (indicated in Figure 1) are approximately 0.2 km s$^{-1}$, allowing these structures to be kinematically distinguished from the envelope emission (with FWHM between 0.6-0.8 km s$^{-1}$).   

\section{Discussion}

HNCO was among the first molecules discovered in the interstellar medium \citep{Snyder71}, that further observations show to be ubiquitous, covering a wide range of objects such as low- and high-mass star formation regions (particularly their envelopes, cores and outflows), and the circumnuclear disk of external galaxies \citep[e.g.,][]{Miettinen2014,Kelly2017,Taniguchi2023,Xie2023}. Low and high angular resolution observations towards massive molecular clouds have demonstrated the existence of extended and compact structures traced by the HNCO emission, with the small scale emission usually associated to cores \citep{Law2021,Taniguchi2023}. The large scale emission, on the other hand, can be related to the densest part of the cloud \citep[see, for example, the maps in Appendix A of ][]{Taniguchi2023} or outflows \citep{Xie2023}. However, given the confusion due to the complex nature of massive regions, it is not possible with sufficient confidence to relate these emission with the filamentary structure of the clouds. Therefore, our observations of L1157 show strong evidence that HNCO is also a tracer of filaments in molecular clouds.

The presence of HNCO in the L1157-mm core/envelope is well-documented, primarily through single-dish pointings. Previous work by \cite{Rodriguez2010} focused exclusively on L1157-mm and the B1/B2 shocks to probe shock-induced chemistry. Consequently, a position isolating the filament emission was not included in their analysis. It was found that indeed at the shock positions this molecule is enhanced by up to almost two orders of magnitude, being the B2 shock the highest HNCO abundance ever recorded. The relative abundance measured towards the protostellar position is 0.3-1.2$\times$10$^{-9}$, while that at the B2 shock is 25-96$\times$10$^{-9}$, with intermediate values of 4.3-17.9$\times$10$^{-9}$ at B1. \cite{Rodriguez2010} noted the similar abundance to the hot corino prototype IRAS 16293, but also comparable to massive star formation regions such as SgrB2M. 

\begin{figure}[!h]
\centering
\includegraphics[angle=0,width=9cm]{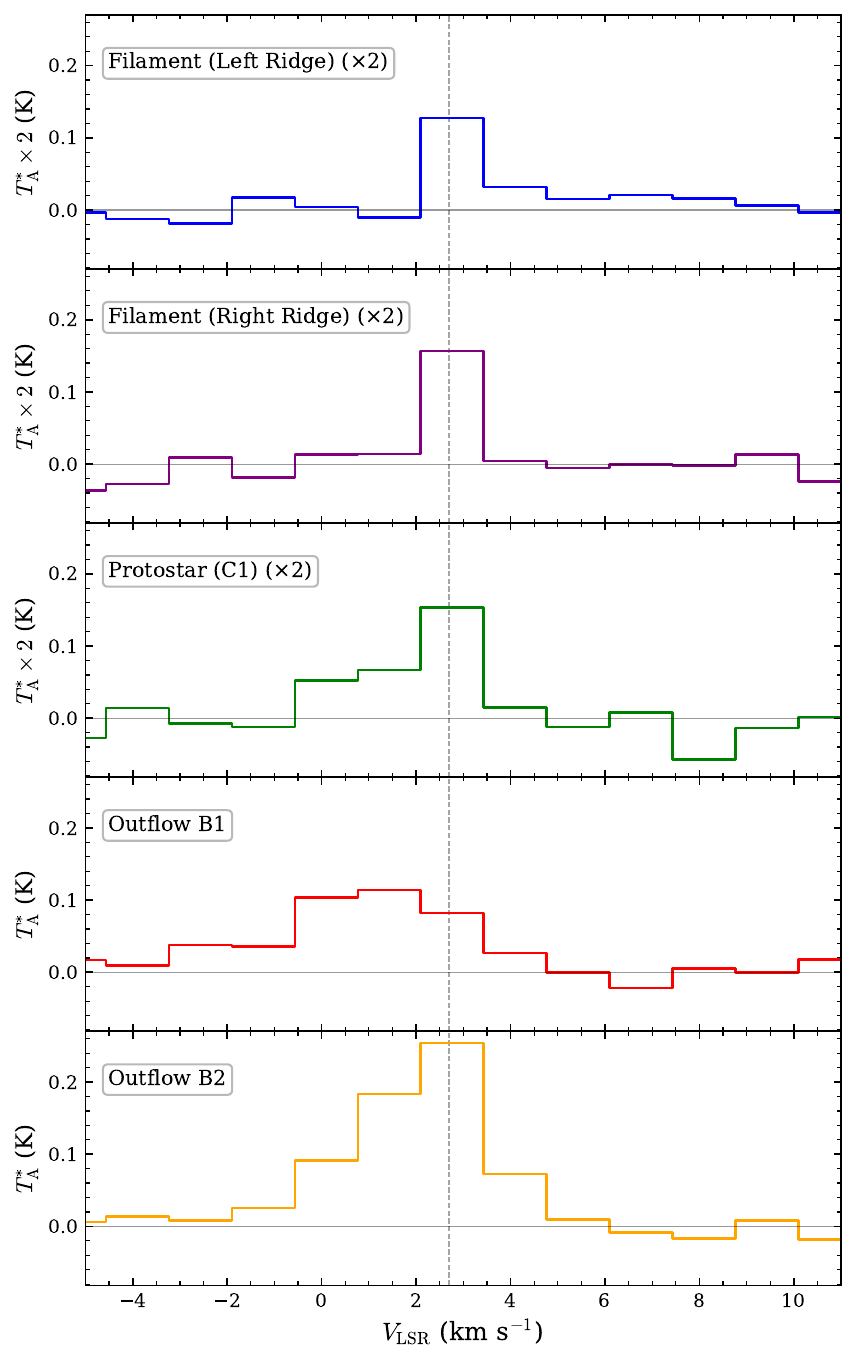}
\caption{HNCO sample spectra at the protostellar, filament and outflow positions indicated in Figure 1.}
\label{spec}
\end{figure}

Previous interferometric observations of HNCO that covered all the blue-lobe and central protostellar regions did not show emission coming from the central protostar \citep{Burkhard2016}, therefore there is no compact emission associated. This together with the fact that our large scale map does not present a clump-like structure towards the protostellar position demonstrate that the detected emission is unlikely tracing a core, but rather the extended emission of the filament itself. Actually, when we compare the line intensities of the newly discovered HNCO filament-associated structures we notice that they have similar peak values to the emission coming from the location of L1157-mm ($\sim$ 80 mK in antenna temperature; see Figure \ref{spec}). Therefore the HNCO column densities should be similar at the location of the protostar and the filamentary large scale structure. The column densities are computed assuming LTE and optically thin emission, 

$N_{\text{tot}} \approx 1.67 \times 10^{14} \, \frac{Q(T_{\text{ex}}) \exp\left(\frac{E_u}{k T_{\text{ex}}}\right)}{\nu_{\text{GHz}} \, (\mu^2 S)_{\text{Debye}^2}} \int T_{\text{mb}} \, dv$

where $\mu^2 S$ is taken from Splatalogue and the rotational constants for the partition function obtained from NIST (approximated as Q $=$ $2 \sqrt{\frac{\pi (k T)^3}{h^3 A B C}}$; with A$=$918.49 GHz, B$=$11.07 GHz, C$=$10.91 GHz). The estimations at the filament positions indicated in Figure 1 are in the range $\sim$1-6$\times$10$^{12}$ cm$^{-2}$, for rotational temperatures of 10-15 K (based on \cite{Rodriguez2010}). \citet{Sharma2020} observed an $N(\text{H}_2)$ contrast of $>20$ between the protostellar region and the filament, with the highest column densities concentrated towards L1157-mm (see the inner versus outer contours in their Figure 3). Because the HNCO line intensity is similar across both regions (as shown in our Figure 3), we expect the relative abundance of HNCO with respect to H$_2$ to be at least an order of magnitude higher in the filament. According to the measurements of \cite{Rodriguez2010} that used multiple HNCO lines to determine accurate excitation temperatures and column densities, the N(HNCO) and X(HNCO) are in the range of 0.2-1.3$\times$10$^{13}$ cm$^{-2}$ and 0.3-1.2$\times$10$^{-9}$, respectively, at the L1157-mm position. Using a column density of 2$\times$10$^{12}$ cm$^{-2}$ (within our estimated range and in agreement with Rodríguez-Fernández et al. lower limit) we therefore obtain at the filament position X(HNCO)$\sim$ 6$\times$10$^{-9}$ \citep[using the H$_2$ column density values from][map]{Sharma2020}. This value is similar to the measurements made towards the Orion Hot Core \citep{Blake1987} and within the range estimated for the B1 shock of the outflow \citep{Rodriguez2010}. Using the LTE column densities (particularly the lower limit of 1$\times$10$^{12}$ cm$^{-2}$ ) and the estimated lower limit of HNCO relative abundance, the combined total mass of both elongations is $\sim$ 0.05 $M_{\odot}$. 

Isocyanic acid is a relatively simple molecule that given its composition consisting of four biogenic elements is thought to be a precursor of more complex organic chemistry with a potential to form molecules such as amino acids, sugars and nucleobases; specifically, HNCO is a peptide bond, --NH--C(=O)--, that resides between any two single amino acid \citep{Gorai2020,Canello2021}.
In star-forming regions there have been a number of investigations that found a correlation of HNCO and NH$_2$CHO \citep[another peptide bond that can link amino acids and form proteins:][]{Gorai2020}, suggesting a chemical link between the two species, either because one forms the other or they are formed from the same precursor \citep{Mendoza2014,Lopez2015,Lopez2019}. The most common mechanism proposed has been the successive hydrogenation of HNCO leading to NH$_2$CHO or two hydrogen abstraction reactions starting from formamide to form HNCO \citep{Quenard2018,Haupa2019,Gorai2020}. 

On the other hand, another modeling perspective suggests that their co-spatial distribution is driven primarily by shared physical conditions (such as temperature and density profiles) rather than a direct chemical link \citep{Quenard2018}. In our observations of the L1157 filament, the estimated HNCO abundance ($\sim$ 6$\times$10$^{-9}$) matches well with theoretical predictions for the early collapse phase (e.g., $\sim$ 6$\times$10$^{-9}$ is reached in about 5.5$\times$10$^4$ yr: see Figure 9 in Gorai et al. 2020). However, unlike the correlated behavior observed in other environments, our data reveal a clear spatial decoupling in the filament: HNCO emission is prominent along the extended quiescent structure, whereas $\mathrm{NH}_2\mathrm{CHO}$ is non-detected in the filament and appears exclusively at the shock positions B1 and B2 (Appendix B; see also López-Sepulcre et al. 2024). This absence of formamide in the cold filamentary gas indicates that HNCO can be efficiently produced in quiescent environments prior to the onset of shock processing or significant warming. Consequently, chemical models must explore alternative formation pathways that account for this spatial decoupling in cold, extended filamentary structures.  

The extended filamentary structure in L1157 represents an early, pristine stage of star formation, as indicated by its cold dust and excitation temperatures ($10\text{--}15\text{ K}$), quiescent kinematics (linewidths $\lesssim 0.2\text{ km s}^{-1}$), and the presence of the chemically young, starless clump C2 alongside the main Class 0 protostar L1157-mm (Sharma et al. 2020). The detection of HNCO across this extended component (well outside the protostellar envelope and shock-heated outflow cavities) demonstrates that HNCO is efficiently produced in the cold, pre-collapse gas prior to the onset of protostellar heating or shock feedback.

\section{Conclusions}

The LMT observations presented here towards the solar-type protostellar system L1157 unambiguously associate the HNCO emission with the filamentary structure from which L1157-mm was formed, and therefore indicates that this molecule is a tracer of densest part of filaments in molecular clouds. The relative abundance estimations (X(HNCO)$\sim$ 6$\times$10$^{-9}$) agree with measurements towards other star-formation regions where this molecule has been detected, but tracing other environments. Although it is possible to reproduce such abundances with existing chemical models, such as the one presented by \citet{Gorai2020}, alternative chemical routes are necessary to account for the observational evidence of the spatial anticorrelation between HNCO and NH$_2$CHO at the studied locations within the filamentary structures. Given its importance in the formation of more complex molecules, the presence of HNCO in the cold, quiescent filamentary gas of L1157 indicates that the building blocks of complex chemistry are already well established in early, pre-collapse stages of star formation, prior to protostellar warming or shock processing.

\begin{acknowledgements}
      This scientific paper uses data obtained from the Large Millimeter Telescope Alfonso Serrano (LMT). The LMT is operated by the Instituto Nacional de Astrofisica, Optica y Electronica (INAOE) and the University of Massachusetts, Amherst (UMass) with long-term binational financial support from the Mexican Secretaria de Ciencias, Humanidades, Tecnologia e Innovacion (SECIHTI, previously CONACYT grant \#297324), and the US National Science Foundation (NSF grant \#2034318) respectively. The SEQUOIA was supported by the National Science Foundation under grant AST-9532046. 
\end{acknowledgements}

\bibliography{bibliog}

@ARTICLE{Arce2008,
       author = {{Arce}, H{\'e}ctor G. and {Santiago-Garc{\'\i}a}, Joaqu{\'\i}n and {J{\o}rgensen}, Jes K. and {Tafalla}, Mario and {Bachiller}, Rafael},
        title = "{Complex Molecules in the L1157 Molecular Outflow}",
      journal = {\apjl},
         year = 2008,
        month = jul,
       volume = {681},
       number = {1},
        pages = {L21},
          doi = {10.1086/590110},
archivePrefix = {arXiv},
       eprint = {0805.2550},
 primaryClass = {astro-ph},
       adsurl = {https://ui.adsabs.harvard.edu/abs/2008ApJ...681L..21A}
}

@ARTICLE{Bachiller2001,
       author = {{Bachiller}, R. and {P{\'e}rez Guti{\'e}rrez}, M. and {Kumar}, M.~S.~N. and {Tafalla}, M.},
        title = "{Chemically active outflow L 1157}",
      journal = {\aap},
         year = 2001,
        month = jun,
       volume = {372},
        pages = {899-912},
          doi = {10.1051/0004-6361:20010519},
       adsurl = {https://ui.adsabs.harvard.edu/abs/2001A&A...372..899B}
}

@ARTICLE{Benedettini2013,
       author = {{Benedettini}, M. and {Viti}, S. and {Codella}, C. and {Gueth}, F. and {G{\'o}mez-Ruiz}, A.~I. and {Bachiller}, R. and {Beltr{\'a}n}, M.~T. and {Busquet}, G. and {Ceccarelli}, C. and {Lefloch}, B.},
        title = "{The B1 shock in the L1157 outflow as seen at high spatial resolution}",
      journal = {\mnras},
         year = 2013,
        month = nov,
       volume = {436},
       number = {1},
        pages = {179-190},
          doi = {10.1093/mnras/stt1559},
archivePrefix = {arXiv},
       eprint = {1309.0433},
 primaryClass = {astro-ph.GA},
       adsurl = {https://ui.adsabs.harvard.edu/abs/2013MNRAS.436..179B}
}

@ARTICLE{Blake1987,
       author = {{Blake}, Geoffrey A. and {Sutton}, E.~C. and {Masson}, C.~R. and {Phillips}, T.~G.},
        title = "{Molecular Abundances in OMC-1: The Chemical Composition of Interstellar Molecular Clouds and the Influence of Massive Star Formation}",
      journal = {\apj},
         year = 1987,
        month = apr,
       volume = {315},
        pages = {621},
          doi = {10.1086/165165},
       adsurl = {https://ui.adsabs.harvard.edu/abs/1987ApJ...315..621B}
}

@ARTICLE{Burkhard2016,
       author = {{Burkhardt}, Andrew M. and {Dollhopf}, Niklaus M. and {Corby}, Joanna F. and {Carroll}, P. Brandon and {Shingledecker}, Christopher N. and {Loomis}, Ryan A. and {Booth}, Shawn Thomas and {Blake}, Geoffrey A. and {Herbst}, Eric and {Remijan}, Anthony J. and {McGuire}, Brett A.},
        title = "{CSO and CARMA Observations of L1157. II. Chemical Complexity in the Shocked Outflow}",
      journal = {\apj},
         year = 2016,
        month = aug,
       volume = {827},
       number = {1},
          eid = {21},
        pages = {21},
          doi = {10.3847/0004-637X/827/1/21},
archivePrefix = {arXiv},
       eprint = {1605.09707},
 primaryClass = {astro-ph.GA},
       adsurl = {https://ui.adsabs.harvard.edu/abs/2016ApJ...827...21B}
}

@ARTICLE{Canello2021,
       author = {{Canelo}, Carla M. and {Bronfman}, Leonardo and {Mendoza}, Edgar and {Duronea}, Nicolas and {Merello}, Manuel and {Carvajal}, Miguel and {Fria{\c{c}}a}, Am{\^a}ncio C.~S. and {Lepine}, Jacques},
        title = "{Isocyanic acid (HNCO) in the hot molecular core G331.512-0.103: observations and chemical modelling}",
      journal = {\mnras},
         year = 2021,
        month = jul,
       volume = {504},
       number = {3},
        pages = {4428-4444},
          doi = {10.1093/mnras/stab1163},
archivePrefix = {arXiv},
       eprint = {2104.10605},
 primaryClass = {astro-ph.GA},
       adsurl = {https://ui.adsabs.harvard.edu/abs/2021MNRAS.504.4428C}
}

@ARTICLE{Chiang10,
       author = {{Chiang}, Hsin-Fang and {Looney}, Leslie W. and {Tobin}, John J. and {Hartmann}, Lee},
        title = "{Probing the Protostellar Envelope Around L1157: The Dust and Gas Connection}",
      journal = {\apj},
         year = 2010,
        month = jan,
       volume = {709},
       number = {1},
        pages = {470-482},
          doi = {10.1088/0004-637X/709/1/470},
archivePrefix = {arXiv},
       eprint = {0912.4527},
 primaryClass = {astro-ph.SR},
       adsurl = {https://ui.adsabs.harvard.edu/abs/2010ApJ...709..470C}
}

@INPROCEEDINGS{Erickson99,
       author = {{Erickson}, N.~R.},
        title = "{Millimeter Wave Heterodyne Receivers and Focal Plane Arrays}",
    booktitle = {Millimeter-Wave Astronomy: Molecular Chemistry \& Physics in Space.},
         year = 1999,
       editor = {{Wall}, W.~F. and {Carrami{\~n}ana}, A. and {Carrasco}, L.},
       series = {Astrophysics and Space Science Library},
       volume = {241},
        month = jan,
        pages = {425},
          doi = {10.1007/978-94-011-4714-9_26},
       adsurl = {https://ui.adsabs.harvard.edu/abs/1999ASSL..241..425E}
}

@ARTICLE{Gorai2020,
       author = {{Gorai}, Prasanta and {Bhat}, Bratati and {Sil}, Milan and {Mondal}, Suman K. and {Ghosh}, Rana and {Chakrabarti}, Sandip K. and {Das}, Ankan},
        title = "{Identification of Prebiotic Molecules Containing Peptide-like Bonds in a Hot Molecular Core, G10.47+0.03}",
      journal = {\apj},
         year = 2020,
        month = jun,
       volume = {895},
       number = {2},
          eid = {86},
        pages = {86},
          doi = {10.3847/1538-4357/ab8871},
archivePrefix = {arXiv},
       eprint = {2003.09188},
 primaryClass = {astro-ph.SR},
       adsurl = {https://ui.adsabs.harvard.edu/abs/2020ApJ...895...86G}
}

@ARTICLE{Gueth1996,
       author = {{Gueth}, F. and {Guilloteau}, S. and {Bachiller}, R.},
        title = "{A precessing jet in the L1157 molecular outflow.}",
      journal = {\aap},
         year = 1996,
        month = mar,
       volume = {307},
        pages = {891-897},
       adsurl = {https://ui.adsabs.harvard.edu/abs/1996A&A...307..891G}
}

@INPROCEEDINGS{Hacar2023,
       author = {{Hacar}, A. and {Clark}, S.~E. and {Heitsch}, F. and {Kainulainen}, J. and {Panopoulou}, G.~V. and {Seifried}, D. and {Smith}, R.},
        title = "{Initial Conditions for Star Formation: a Physical Description of the Filamentary ISM}",
    booktitle = {Protostars and Planets VII},
         year = 2023,
       editor = {{Inutsuka}, S. and {Aikawa}, Y. and {Muto}, T. and {Tomida}, K. and {Tamura}, M.},
       series = {Astronomical Society of the Pacific Conference Series},
       volume = {534},
        month = jul,
        pages = {153},
          doi = {10.48550/arXiv.2203.09562},
archivePrefix = {arXiv},
       eprint = {2203.09562},
 primaryClass = {astro-ph.GA},
       adsurl = {https://ui.adsabs.harvard.edu/abs/2023ASPC..534..153H}
}

@ARTICLE{Haupa2019,
       author = {{Haupa}, Karolina A. and {Tarczay}, Gy{\"o}rgy and {Lee}, Yuan-Pern},
        title = "{Hydrogen Abstraction/Addition Tunneling Reactions Elucidate the Interstellar H2NCHO/HNCO Ratio and H2 Formation}",
      journal = {Journal of the American Chemical Society},
         year = 2019,
        month = jul,
       volume = {141},
       number = {29},
        pages = {11614-11620},
          doi = {10.1021/jacs.9b04491},
       adsurl = {https://ui.adsabs.harvard.edu/abs/2019JAChS.14111614H}
}

@INPROCEEDINGS{Hughes2020,
       author = {{Hughes}, David H. and {Schloerb}, F. Peter and {Aretxaga}, Itziar and {Castillo-Dom{\'\i}nguez}, Edgar and {Ch{\'a}vez Dagostino}, Miguel and {Col{\'\i}n}, Edgar and {Erickson}, Neal and {Ferrusca Rodriguez}, Daniel and {Gale}, David M. and {G{\'o}mez-Ruiz}, Arturo and {Hern{\'a}ndez Rebollar}, Jos{\'e} Luis and {Heyer}, Mark and {Lowenthal}, James and {Monta{\~n}a}, Alfredo and {Moreno Nolasco}, Marcos Emir and {Narayanan}, Gopal and {Pope}, Alexandra and {Rodr{\'\i}guez-Montoya}, Iv{\'a}n. and {S{\'a}nchez-Arg{\"u}elles}, David and {Smith}, David and {Souccar}, Kamal and {de la Rosa Becerra}, Miguel Vel{\'a}zquez and {Wilson}, Grant W. and {Yun}, Min S.},
        title = "{The Large Millimeter Telescope (LMT) Alfonso Serrano: current status and telescope performance}",
    booktitle = {Ground-based and Airborne Telescopes VIII},
         year = 2020,
       editor = {{Marshall}, Heather K. and {Spyromilio}, Jason and {Usuda}, Tomonori},
       series = {Society of Photo-Optical Instrumentation Engineers (SPIE) Conference Series},
       volume = {11445},
        month = dec,
          eid = {1144522},
        pages = {1144522},
          doi = {10.1117/12.2561893},
       adsurl = {https://ui.adsabs.harvard.edu/abs/2020SPIE11445E..22H}
}

@ARTICLE{Kelly2017,
       author = {{Kelly}, G. and {Viti}, S. and {Garc{\'\i}a-Burillo}, S. and {Fuente}, A. and {Usero}, A. and {Krips}, M. and {Neri}, R.},
        title = "{Molecular shock tracers in NGC 1068: SiO and HNCO}",
      journal = {\aap},
         year = 2017,
        month = jan,
       volume = {597},
          eid = {A11},
        pages = {A11},
          doi = {10.1051/0004-6361/201628946},
archivePrefix = {arXiv},
       eprint = {1609.02023},
 primaryClass = {astro-ph.GA},
       adsurl = {https://ui.adsabs.harvard.edu/abs/2017A&A...597A..11K}
}

@ARTICLE{Law2021,
       author = {{Law}, Charles J. and {Zhang}, Qizhou and {{\"O}berg}, Karin I. and {Galv{\'a}n-Madrid}, Roberto and {Keto}, Eric and {Liu}, Hauyu Baobab and {Ho}, Paul T.~P.},
        title = "{Subarcsecond Imaging of the Complex Organic Chemistry in Massive Star-forming Region G10.6-0.4}",
      journal = {\apj},
         year = 2021,
        month = mar,
       volume = {909},
       number = {2},
          eid = {214},
        pages = {214},
          doi = {10.3847/1538-4357/abdeb8},
archivePrefix = {arXiv},
       eprint = {2101.07801},
 primaryClass = {astro-ph.GA},
       adsurl = {https://ui.adsabs.harvard.edu/abs/2021ApJ...909..214L}
}

@ARTICLE{Lefloch2017,
       author = {{Lefloch}, Bertrand and {Ceccarelli}, C. and {Codella}, C. and {Favre}, C. and {Podio}, L. and {Vastel}, C. and {Viti}, S. and {Bachiller}, R.},
        title = "{L1157-B1, a factory of complex organic molecules in a solar-type star-forming region}",
      journal = {\mnras},
         year = 2017,
        month = jul,
       volume = {469},
       number = {1},
        pages = {L73-L77},
          doi = {10.1093/mnrasl/slx050},
archivePrefix = {arXiv},
       eprint = {1704.04646},
 primaryClass = {astro-ph.GA},
       adsurl = {https://ui.adsabs.harvard.edu/abs/2017MNRAS.469L..73L}
}

@ARTICLE{Lopez2015,
       author = {{L{\'o}pez-Sepulcre}, A. and {Jaber}, Ali A. and {Mendoza}, E. and {Lefloch}, B. and {Ceccarelli}, C. and {Vastel}, C. and {Bachiller}, R. and {Cernicharo}, J. and {Codella}, C. and {Kahane}, C. and {Kama}, M. and {Tafalla}, M.},
        title = "{Shedding light on the formation of the pre-biotic molecule formamide with ASAI}",
      journal = {\mnras},
         year = 2015,
        month = may,
       volume = {449},
       number = {3},
        pages = {2438-2458},
          doi = {10.1093/mnras/stv377},
archivePrefix = {arXiv},
       eprint = {1502.05762},
 primaryClass = {astro-ph.SR},
       adsurl = {https://ui.adsabs.harvard.edu/abs/2015MNRAS.449.2438L}
}

@ARTICLE{Lopez2019,
       author = {{L{\'o}pez-Sepulcre}, Ana and {Balucani}, Nadia and {Ceccarelli}, Cecilia and {Codella}, Claudio and {Dulieu}, Fran{\c{c}}ois and {Theul{\'e}}, Patrice},
        title = "{Interstellar Formamide (NH2CHO), a Key Prebiotic Precursor}",
      journal = {ACS Earth and Space Chemistry},
         year = 2019,
        month = oct,
       volume = {3},
       number = {10},
        pages = {2122-2137},
          doi = {10.1021/acsearthspacechem.9b00154},
archivePrefix = {arXiv},
       eprint = {1909.11770},
 primaryClass = {astro-ph.SR},
       adsurl = {https://ui.adsabs.harvard.edu/abs/2019ESC.....3.2122L}
}

@ARTICLE{Lopez2024,
       author = {{L{\'o}pez-Sepulcre}, A. and {Codella}, C. and {Ceccarelli}, C. and {Podio}, L. and {Robuschi}, J.},
        title = "{Full interferometric map of the L1157 southern outflow: Formamide (NH$_{2}$CHO) can form in the gas, after all}",
      journal = {\aap},
         year = 2024,
        month = dec,
       volume = {692},
          eid = {A120},
        pages = {A120},
          doi = {10.1051/0004-6361/202450169},
archivePrefix = {arXiv},
       eprint = {2411.00495},
 primaryClass = {astro-ph.GA},
       adsurl = {https://ui.adsabs.harvard.edu/abs/2024A&A...692A.120L}
}

@ARTICLE{Mendoza2014,
       author = {{Mendoza}, Edgar and {Lefloch}, B. and {L{\'o}pez-Sepulcre}, A. and {Ceccarelli}, C. and {Codella}, C. and {Boechat-Roberty}, H.~M. and {Bachiller}, R.},
        title = "{Molecules with a peptide link in protostellar shocks: a comprehensive study of L1157}",
      journal = {\mnras},
         year = 2014,
        month = nov,
       volume = {445},
       number = {1},
        pages = {151-161},
          doi = {10.1093/mnras/stu1718},
archivePrefix = {arXiv},
       eprint = {1408.4857},
 primaryClass = {astro-ph.SR},
       adsurl = {https://ui.adsabs.harvard.edu/abs/2014MNRAS.445..151M}
}

@ARTICLE{Miettinen2014,
       author = {{Miettinen}, O.},
        title = "{A MALT90 study of the chemical properties of massive clumps and filaments of infrared dark clouds}",
      journal = {\aap},
         year = 2014,
        month = feb,
       volume = {562},
          eid = {A3},
        pages = {A3},
          doi = {10.1051/0004-6361/201322596},
archivePrefix = {arXiv},
       eprint = {1311.4300},
 primaryClass = {astro-ph.GA},
       adsurl = {https://ui.adsabs.harvard.edu/abs/2014A&A...562A...3M}
}

@ARTICLE{Muller2005,
       author = {{M{\"u}ller}, Holger S.~P. and {Schl{\"o}der}, Frank and {Stutzki}, J{\"u}rgen and {Winnewisser}, Gisbert},
        title = "{The Cologne Database for Molecular Spectroscopy, CDMS: a useful tool for astronomers and spectroscopists}",
      journal = {Journal of Molecular Structure},
         year = 2005,
        month = may,
       volume = {742},
       number = {1-3},
        pages = {215-227},
          doi = {10.1016/j.molstruc.2005.01.027},
       adsurl = {https://ui.adsabs.harvard.edu/abs/2005JMoSt.742..215M}
}

@ARTICLE{Nisini2010,
       author = {{Nisini}, Brunella and {Giannini}, Teresa and {Neufeld}, David A. and {Yuan}, Yuan and {Antoniucci}, Simone and {Bergin}, Edwin A. and {Melnick}, Gary J.},
        title = "{Spitzer Spectral Line Mapping of Protostellar Outflows. II. H$_{2}$ Emission in L1157}",
      journal = {\apj},
         year = 2010,
        month = nov,
       volume = {724},
       number = {1},
        pages = {69-79},
          doi = {10.1088/0004-637X/724/1/69},
archivePrefix = {arXiv},
       eprint = {1009.3834},
 primaryClass = {astro-ph.SR},
       adsurl = {https://ui.adsabs.harvard.edu/abs/2010ApJ...724...69N}
}

@INPROCEEDINGS{Pineda2023,
       author = {{Pineda}, J.~E. and {Arzoumanian}, D. and {Andre}, P. and {Friesen}, R.~K. and {Zavagno}, A. and {Clarke}, S.~D. and {Inoue}, T. and {Chen}, C. and {Lee}, Y. and {Soler}, J.~D. and {Kuffmeier}, M.},
        title = "{From Bubbles and Filaments to Cores and Disks: Gas Gathering and Growth of Structure Leading to the Formation of Stellar Systems}",
    booktitle = {Protostars and Planets VII},
         year = 2023,
       editor = {{Inutsuka}, S. and {Aikawa}, Y. and {Muto}, T. and {Tomida}, K. and {Tamura}, M.},
       series = {Astronomical Society of the Pacific Conference Series},
       volume = {534},
        month = jul,
        pages = {233},
          doi = {10.48550/arXiv.2205.03935},
archivePrefix = {arXiv},
       eprint = {2205.03935},
 primaryClass = {astro-ph.GA},
       adsurl = {https://ui.adsabs.harvard.edu/abs/2023ASPC..534..233P}
}

@ARTICLE{Podio2016,
       author = {{Podio}, L. and {Codella}, C. and {Gueth}, F. and {Cabrit}, S. and {Maury}, A. and {Tabone}, B. and {Lef{\`e}vre}, C. and {Anderl}, S. and {Andr{\'e}}, P. and {Belloche}, A. and {Bontemps}, S. and {Hennebelle}, P. and {Lefloch}, B. and {Maret}, S. and {Testi}, L.},
        title = "{First image of the L1157 molecular jet by the CALYPSO IRAM-PdBI survey}",
      journal = {\aap},
         year = 2016,
        month = sep,
       volume = {593},
          eid = {L4},
        pages = {L4},
          doi = {10.1051/0004-6361/201628876},
archivePrefix = {arXiv},
       eprint = {1608.05026},
 primaryClass = {astro-ph.GA},
       adsurl = {https://ui.adsabs.harvard.edu/abs/2016A&A...593L...4P}
}

@ARTICLE{Priestley2023,
       author = {{Priestley}, F.~D. and {Arzoumanian}, D. and {Whitworth}, A.~P.},
        title = "{Line emission from filaments in molecular clouds}",
      journal = {\mnras},
         year = 2023,
        month = jul,
       volume = {522},
       number = {3},
        pages = {3890-3897},
          doi = {10.1093/mnras/stad1229},
archivePrefix = {arXiv},
       eprint = {2304.12357},
 primaryClass = {astro-ph.GA},
       adsurl = {https://ui.adsabs.harvard.edu/abs/2023MNRAS.522.3890P}
}

@ARTICLE{Quenard2018,
       author = {{Qu{\'e}nard}, David and {Jim{\'e}nez-Serra}, Izaskun and {Viti}, Serena and {Holdship}, Jonathan and {Coutens}, Audrey},
        title = "{Chemical modelling of complex organic molecules with peptide-like bonds in star-forming regions}",
      journal = {\mnras},
         year = 2018,
        month = feb,
       volume = {474},
       number = {2},
        pages = {2796-2812},
          doi = {10.1093/mnras/stx2960},
archivePrefix = {arXiv},
       eprint = {1711.05184},
 primaryClass = {astro-ph.SR},
       adsurl = {https://ui.adsabs.harvard.edu/abs/2018MNRAS.474.2796Q}
}

@ARTICLE{Rodriguez2010,
       author = {{Rodr{\'\i}guez-Fern{\'a}ndez}, N.~J. and {Tafalla}, M. and {Gueth}, F. and {Bachiller}, R.},
        title = "{HNCO enhancement by shocks in the L1157 molecular outflow}",
      journal = {\aap},
         year = 2010,
        month = jun,
       volume = {516},
          eid = {A98},
        pages = {A98},
          doi = {10.1051/0004-6361/201013997},
archivePrefix = {arXiv},
       eprint = {1003.4219},
 primaryClass = {astro-ph.GA},
       adsurl = {https://ui.adsabs.harvard.edu/abs/2010A&A...516A..98R}
}

@ARTICLE{Sharma2020,
       author = {{Sharma}, Ekta and {Gopinathan}, Maheswar and {Soam}, Archana and {Lee}, Chang Won and {Kim}, Shinyoung and {Ghosh}, Tuhin and {Tej}, Anandmayee and {Kim}, Gwanjeong and {Sharma}, Neha and {Saha}, Piyali},
        title = "{Distance, magnetic field, and kinematics of the filamentary cloud LDN 1157}",
      journal = {\aap},
         year = 2020,
        month = jul,
       volume = {639},
          eid = {A133},
        pages = {A133},
          doi = {10.1051/0004-6361/202037438},
archivePrefix = {arXiv},
       eprint = {2005.02781},
 primaryClass = {astro-ph.SR},
       adsurl = {https://ui.adsabs.harvard.edu/abs/2020A&A...639A.133S}
}

@INPROCEEDINGS{Snyder71,
       author = {{Snyder}, L.~E. and {Bhul}, D.},
        title = "{Detection of Interstellar Isocyanic Acid, Methylacetylene and Hydrogen Isocyanide.}",
    booktitle = {Bulletin of the American Astronomical Society},
         year = 1971,
       volume = {3},
        month = jun,
        pages = {388},
       adsurl = {https://ui.adsabs.harvard.edu/abs/1971BAAS....3Q.388S}
}

@ARTICLE{Taniguchi2023,
       author = {{Taniguchi}, Kotomi and {Sanhueza}, Patricio and {Olguin}, Fernando A. and {Gorai}, Prasanta and {Das}, Ankan and {Nakamura}, Fumitaka and {Saito}, Masao and {Zhang}, Qizhou and {Lu}, Xing and {Li}, Shanghuo and {Chen}, Huei-Ru Vivien},
        title = "{Digging into the Interior of Hot Cores with the ALMA (DIHCA). III. The Chemical Link between NH$_{2}$CHO, HNCO, and H$_{2}$CO}",
      journal = {\apj},
         year = 2023,
        month = jun,
       volume = {950},
       number = {1},
          eid = {57},
        pages = {57},
          doi = {10.3847/1538-4357/acca1d},
archivePrefix = {arXiv},
       eprint = {2304.00267},
 primaryClass = {astro-ph.GA},
       adsurl = {https://ui.adsabs.harvard.edu/abs/2023ApJ...950...57T}
}

@ARTICLE{Tobin2022,
       author = {{Tobin}, John J. and {Cox}, Erin G. and {Looney}, Leslie W.},
        title = "{A 16 au Binary in the Class 0 Protostar L1157 MMS}",
      journal = {\apj},
         year = 2022,
        month = mar,
       volume = {928},
       number = {1},
          eid = {61},
        pages = {61},
          doi = {10.3847/1538-4357/ac5594},
archivePrefix = {arXiv},
       eprint = {2202.07667},
 primaryClass = {astro-ph.SR},
       adsurl = {https://ui.adsabs.harvard.edu/abs/2022ApJ...928...61T}
}

@ARTICLE{Tobin2011,
       author = {{Tobin}, John J. and {Hartmann}, Lee and {Chiang}, Hsin-Fang and {Looney}, Leslie W. and {Bergin}, Edwin A. and {Chandler}, Claire J. and {Masqu{\'e}}, Josep M. and {Maret}, S{\'e}bastien and {Heitsch}, Fabian},
        title = "{Complex Structure in Class 0 Protostellar Envelopes. II. Kinematic Structure from Single-dish and Interferometric Molecular Line Mapping}",
      journal = {\apj},
         year = 2011,
        month = oct,
       volume = {740},
       number = {1},
          eid = {45},
        pages = {45},
          doi = {10.1088/0004-637X/740/1/45},
archivePrefix = {arXiv},
       eprint = {1107.4361},
 primaryClass = {astro-ph.SR},
       adsurl = {https://ui.adsabs.harvard.edu/abs/2011ApJ...740...45T}
}

@ARTICLE{Xie2023,
       author = {{Xie}, Jinjin and {Li}, Juan and {Wang}, Junzhi and {Liu}, Shu and {Yang}, Kai and {Quan}, Donghui and {Zheng}, Siqi and {Li}, Yuqiang and {Wu}, Jingwen and {Duan}, Yan and {Li}, Di},
        title = "{Imaging Molecular Outflow in Massive Star-forming Regions with HNCO Lines}",
      journal = {\apj},
         year = 2023,
        month = jun,
       volume = {949},
       number = {2},
          eid = {89},
        pages = {89},
          doi = {10.3847/1538-4357/acc83f},
archivePrefix = {arXiv},
       eprint = {2303.14866},
 primaryClass = {astro-ph.GA},
       adsurl = {https://ui.adsabs.harvard.edu/abs/2023ApJ...949...89X}
}

@ARTICLE{Yamaguchi2012,
       author = {{Yamaguchi}, Takahiro and {Takano}, Shuro and {Watanabe}, Yoshimasa and {Sakai}, Nami and {Sakai}, Takeshi and {Liu}, Sheng-Yuan and {Su}, Yu-Nung and {Hirano}, Naomi and {Takakuwa}, Shigehisa and {Aikawa}, Yuri and {Nomura}, Hideko and {Yamamoto}, Satoshi},
        title = "{The 3 mm Spectral Line Survey toward the Lynds 1157 B1 Shocked Region. I. Data}",
      journal = {\pasj},
         year = 2012,
        month = oct,
       volume = {64},
          eid = {105},
        pages = {105},
          doi = {10.1093/pasj/64.5.105},
       adsurl = {https://ui.adsabs.harvard.edu/abs/2012PASJ...64..105Y}
}

@ARTICLE{Zhang2000,
       author = {{Zhang}, Q. and {Ho}, P.~T.~P. and {Wright}, M.~C.~H.},
        title = "{The SIO and CS Emission in the Molecular Outflow toward L1157}",
      journal = {\aj},
         year = 2000,
        month = mar,
       volume = {119},
       number = {3},
        pages = {1345-1351},
          doi = {10.1086/301274},
       adsurl = {https://ui.adsabs.harvard.edu/abs/2000AJ....119.1345Z}
}

@ARTICLE{Zucker2019,
       author = {{Zucker}, Catherine and {Speagle}, Joshua S. and {Schlafly}, Edward F. and {Green}, Gregory M. and {Finkbeiner}, Douglas P. and {Goodman}, Alyssa A. and {Alves}, Jo{\~a}o},
        title = "{A Large Catalog of Accurate Distances to Local Molecular Clouds: The Gaia DR2 Edition}",
      journal = {\apj},
         year = 2019,
        month = jul,
       volume = {879},
       number = {2},
          eid = {125},
        pages = {125},
          doi = {10.3847/1538-4357/ab2388},
archivePrefix = {arXiv},
       eprint = {1902.01425},
 primaryClass = {astro-ph.GA},
       adsurl = {https://ui.adsabs.harvard.edu/abs/2019ApJ...879..125Z}
}

\begin{appendix}
\onecolumn
\nolinenumbers

\section{All moments maps of the HNCO emission}

\begin{figure*}[!ht]
\centering
\includegraphics[angle=0,width=15cm]{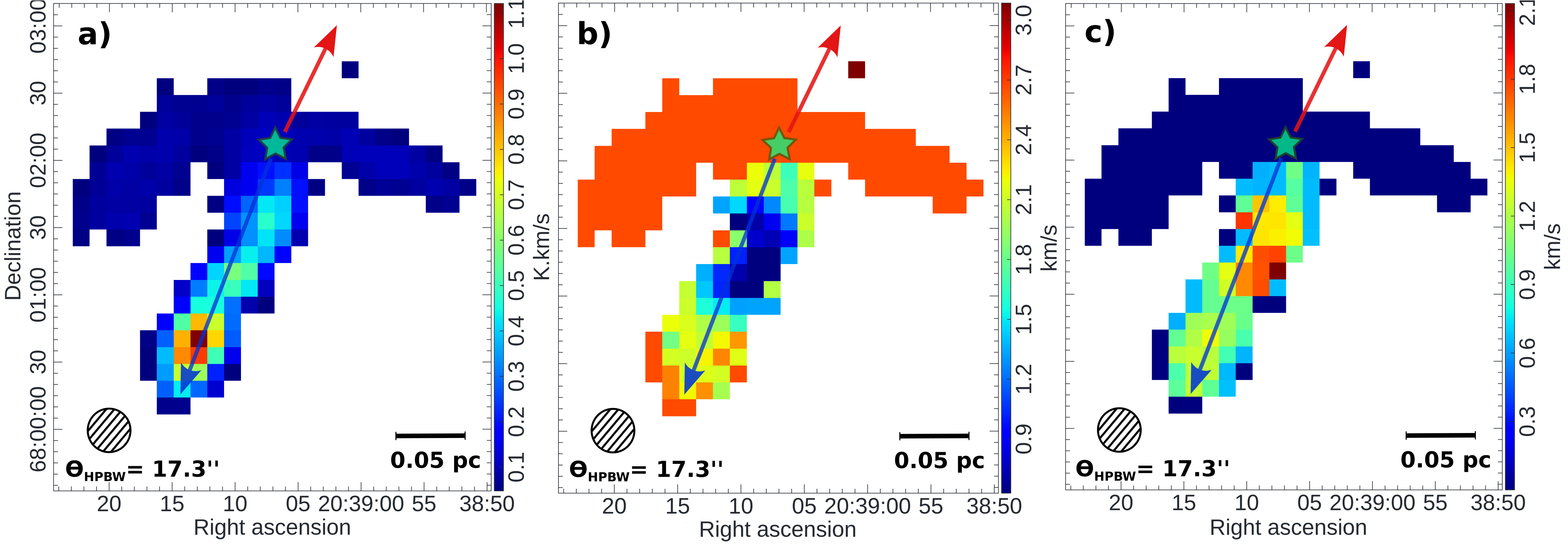}
\caption{Integrated intensity (moment 0; a), velocity field (moment 1; b) and velocity dispersion (moment 2; c) maps of the HNCO (4$_{0,4}$--3$_{0,3}$) emission. The star symbol and the arrows indicate the protostellar position (L1157-mm) and the blue/red outflow lobes, respectively. The beam size and a physical scale bar are also shown.}
\label{moments}
\end{figure*}

An overall view of the HNCO (4$_{0,4}$--3$_{0,3}$) emission can be obtained from the moment maps presented in Figure \ref{moments}. First of all, one can notice, based on the integrated intensity emission (moment 0), that the emission is dominated by the outflow component. However, unfortunately, given the limited spectral resolution of 1.3 km s$^{-1}$, the emission from the filament and the envelope/core remains blended within a single channel ($V_{lsr} = 2.7$ km s$^{-1}$), preventing a clear kinematic decomposition of these structures in the moment 1 and 2 maps, from which we can only distinguish between the low and high velocity components of the outflow and the elongations related to the filamentary structure. 

\section{NH$_2$CHO integrated intensity map and sample spectra}

\begin{figure}[!h]
\centering
    \begin{minipage}{.5\textwidth}
        \centering
        \includegraphics[width=7cm]{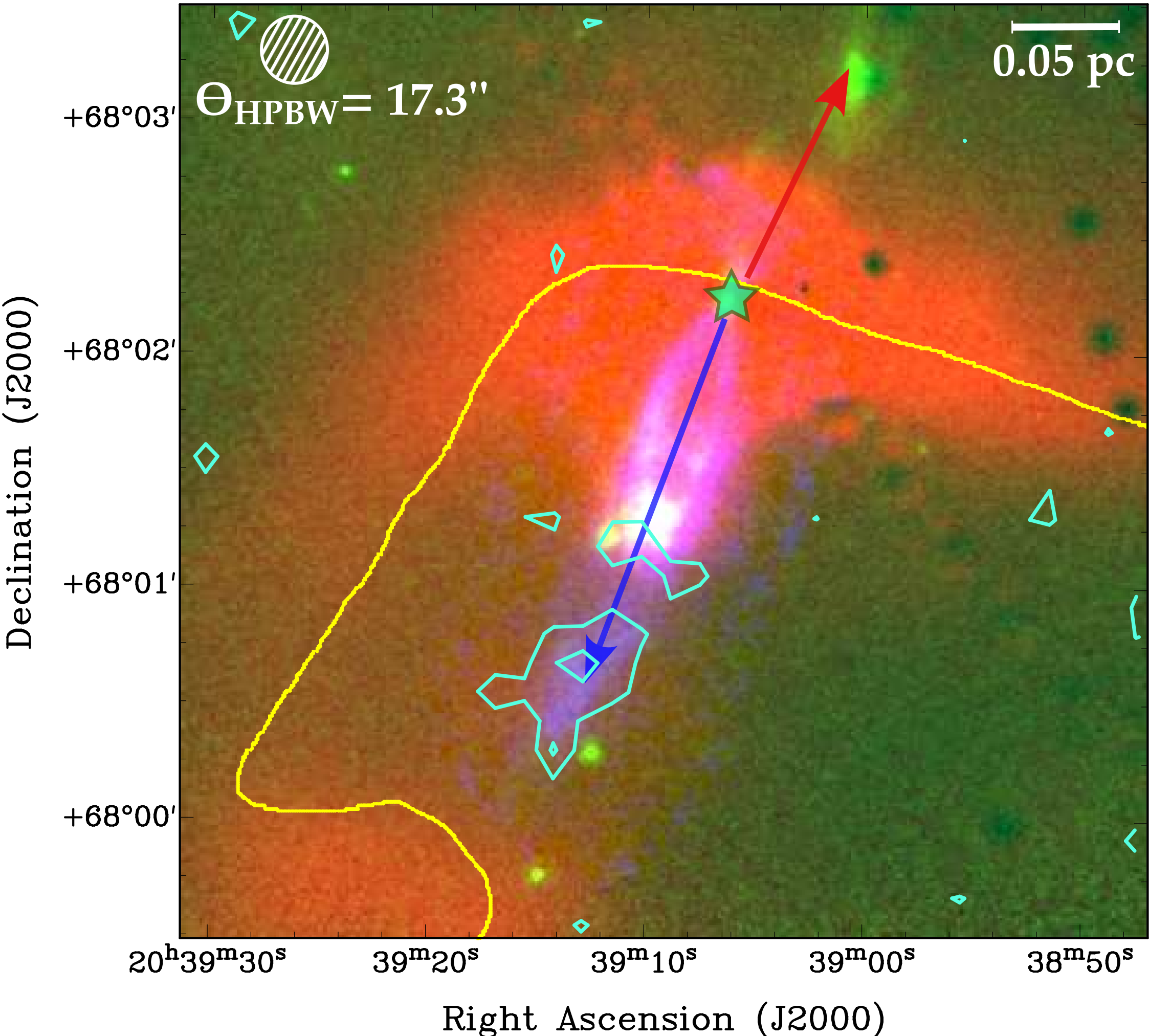}
        \caption{NH$_2$CHO (4$_{1,3}$--3$_{1,2}$) integrated emission towards the blue lobe of L1157 outflow (contours). The background image consists of the 250 micron emission from \cite{Sharma2020} (red), 8 micron emission from Spitzer (green) and the CO (1--0) integrated emission from \cite{Gueth1996} (blue). The star symbol and the arrows indicate the protostellar position (L1157-mm) and the blue/red outflow lobes, respectively. The beam size and a physical scale bar are also shown.}
        \label{NH2CHO}
    \end{minipage}%
    \begin{minipage}{0.5\textwidth}
        \centering
        \includegraphics[width=9cm]{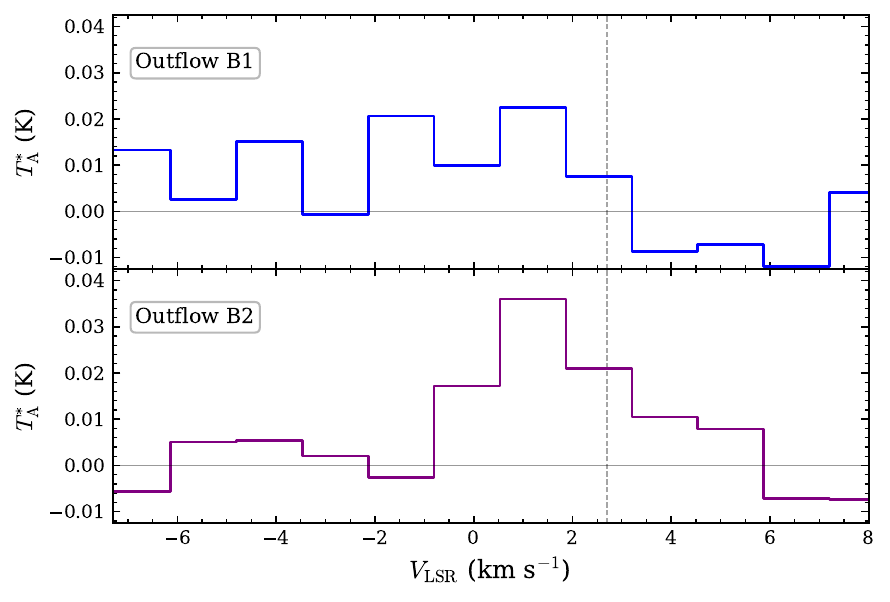}
        \caption{NH$_2$CHO (4$_{1,3}$--3$_{1,2}$) sample spectra at the emission peaks in Figure \ref{NH2CHO}.}
        \label{NH2CHO-spec}
    \end{minipage}
\end{figure}

Figure \ref{NH2CHO} shows the total NH$_2$CHO integrated emission, which we detected in only a couple of channels around the cloud velocity. The brightness distribution is consistent with the interferometric results presented by \citet{Lopez2024}, in which the emission comes from the B1 and B2 shocks, with the latter showing the strongest formamide emission. Figure \ref{NH2CHO-spec} present sample spectra at the NH$_2$CHO emission peaks.

\end{appendix}

\end{document}